\documentstyle[multicol,prl,aps,graphicx]{revtex}
\setkeys{Gin}{width=0.9\linewidth}

\begin{document}
\draft

\title{Paramagnetic reentrant effect in high purity mesoscopic AgNb proximity structures}

\author{F. Bernd M{\"u}ller-Allinger and Ana Celia Mota}
\address{Laboratorium f{\"u}r Festk{\"o}rperphysik, Eidgen{\"o}ssische 
         Technische Hochschule Z{\"u}rich, \\ 
         8093 Z{\"u}rich, Switzerland}

\date{\today}
\maketitle

\begin{abstract}
We discuss the magnetic response of clean Ag coated Nb proximity 
cylinders in the temperature range $150\,\mu{\mathrm K}<T<9\,{\mathrm 
K}$.  In the mesoscopic temperature regime, the normal 
metal-superconductor system shows the yet unexplained paramagnetic 
reentrant effect, discovered some years ago [P. Visani, A. C. Mota, 
and A. Pollini, Phys.  Rev.  Lett.  \textbf{65}, 1514 (1990)], 
superimposing on full Meissner screening.  The logarithmic slope of 
the reentrant paramagnetic susceptibility $\chi_{\mathrm 
para}(T)\propto\exp{(-L/\xi_{N})}$ is limited by the condition 
$\xi_{N}=n\, L$, with $\xi_{N}=\hbar v_{F}/2\pi k_{B}T$, the thermal 
coherence length and $n=1,2,4$.  At the lowest temperatures, 
$\chi_{\mathrm para}$ compensates the diamagnetic susceptibility of 
the \textit{whole} AgNb structure.
\end{abstract}

\begin{multicols}{2}
\narrowtext
In recent years, there has been extensive experimental and theoretical 
work in the field of mesoscopic systems, including superconducting 
structures in proximity with normal metals \cite{review}.  In 
particular, the paramagnetic reentrance phenomenon \cite{visaniprl} 
has received a wide interest, mainly because of being promoted by the 
recent understanding of the high-temperature diamagnetic response of 
rather clean normal-metal--superconductor (NS) proximity structures 
\cite{bmueller1} in the context of the quasiclassical Eilenberger 
theory including elastic scattering \cite{belzig}.  In this Letter, we 
discuss the very low temperature reentrant behavior of two of the AgNb 
samples of Ref.\ \cite{bmueller1}, covering a larger mesoscopic regime 
with respect to previous measurements \cite{visaniprl}.

Recently, two Letters \cite{bruder,fauchere99} have addressed the 
origin of paramagnetic currents in NS systems, which might lead to an 
understanding of the paramagnetic reentrance phenomenon.  The work of 
Bruder and Imry \cite{bruder} is based on the presence of 
non-Andreev-reflecting semiclassical trajectories at the outer surface 
of a nonsingly connected proximity system (glancing states), which 
carry predominantly paramagnetic currents.  This work has been subject 
to debate because of the small magnitude \cite{comment}.  A different, 
more elaborate approach by Fauch\`ere \textit{et al}.  
\cite{fauchere99} assumes a net repulsive interaction in the noble 
metals.  The $\pi$ shift of the order parameter at the NS interface 
then leads to a paramagnetic instability of Andreev pairs.

The first work reflects the cylindrical geometry of our NS system, but 
it does not address the experimental signatures of the paramagnetic 
reentrance, namely absolute value of order 1, temperature dependence, 
nonlinearity, hysteresis, or dissipation \cite{visaniprl}.  The latter 
three features might give evidence for a spontaneous magnetization in 
the samples, as proposed in Ref.\ \cite{fauchere99}.  However, the 
second theoretical approach \cite{fauchere99} does not reach beyond 
qualitative accordance with our experiment.

Here we discuss an investigation of the paramagnetic reentrant effect 
extending to the $\mu$K region.  By covering five decades in 
temperature, we have been able to extract the correct temperature 
dependence of the NS proximity structure below $T_{c}$.  Over a large 
mesoscopic regime, the cylindrical structure clearly displays 
different levels of coherence along integer multiples of the wire 
perimeter $L$.  Unfortunately, neither of the two present theories 
\cite{bruder,fauchere99}, which discuss different sources of 
paramagnetism in NS structures, obtain the correct absolute value as 
well as the $T$ dependence of the paramagnetic reentrant effect.

The two samples reported here are ensembles of cylindrical wires with a 
superconducting core of soft niobium ($RRR\approx 300$) concentrically 
embedded in a normal-metal matrix of 6N silver. 
Their total diameter was mechanically reduced by several steps of 
swagging and co-drawing \cite{flukiger} to final values 
$41\,\mu{\mathrm m}$ and $23\,\mu{\mathrm m}$, with normal layer thicknesses 
$d_{N}=5.5\,\mu{\mathrm m}$ and $3.3\,\mu{\mathrm m}$ (the ratio 
$d_{N}/L$ is approximately the same), respectively. 
The samples were 
\begin{figure}
\includegraphics[width=0.96\linewidth]{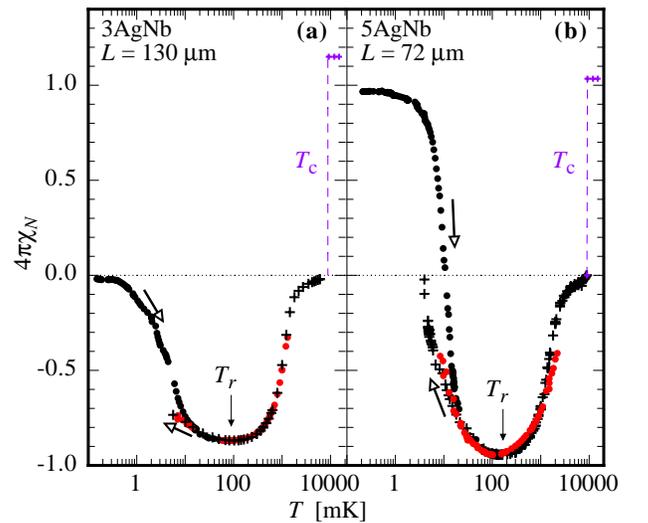}
	\caption{Magnetic susceptibility $\chi_{N}(T)$, between 
	$150\,\mu{\mathrm K}$ and $9\,{\mathrm K}$.  For both samples we 
	show $\chi_{ac}(T)$ ($+$) and $\chi_{dc}(T)$ ($\bullet$).  The 
	arrows mark the direction of $T$--changes.  }
    \protect\label{Xdc(logT)}
\end{figure}

\noindent annealed after the last drawing, 
and values of the mean free path $\ell_{N}\sim(0.5$--$0.8)d_{N}$ 
were obtained.  For more details see 
Refs.\ \cite{journlowtempphys,bmueller1}.

Extensions of the measurements to $\mu$K temperatures were performed 
at the ultralow temperature (ULT) facility at the University of 
Bayreuth.  There, an experimental setup was installed for inductive 
measurements using an rf--SQUID sensor.  Magnetic fields were applied 
along the axis of the wires.  
%

For the ULT experiments, we took parts of the wire bundle measured in 
our dilution refrigerator \cite{visaniprl,bmueller1}, and glued them 
with GE 7031 varnish to high purity gold foils tightly attached to a 
silver finger, in good electrical contact with the Cu demagnetization 
stage \cite{gloos88}.  Thus, about 200 wires were mounted.  
Temperatures were measured with a Pt pulsed NMR thermometer 
\cite{gloos88}.

In the following we report on the temperature dependent magnetic 
susceptibility of our relatively clean silver-niobium samples 3AgNb 
[$d_{N}=5.5\,\mu{\mathrm m}$] and 5AgNb [$d_{N}=3.3\,\mu{\mathrm m}$].  
In Fig.\ \ref{Xdc(logT)} the total magnetic susceptibility is shown as 
a function of temperature.  We show $\chi_{ac}(T)$ between 
$4\,{\mathrm mK}$ and $9\,{\mathrm K}$, measured in our dilution 
refrigerator with field amplitude $H_{ac}=33\,{\mathrm mOe}$ and 
frequency $\nu=80\,{\mathrm Hz}$, as well as $\chi_{dc}(T)$ at 
constant $H_{dc}$ at ULT and LT.

At temperatures below the critical temperature of Nb 
$T_{c}=9.2\,{\mathrm K}$, the magnetic susceptibility of the N layer 
exhibits diamagnetism induced through Andreev reflection at the highly 
transparent NS interface.  At lower temperatures, it develops almost 
total Meissner screening in the Ag layer \cite{bmueller1}.  Below 
$T_{r}\sim 100\,{\mathrm mK}$ the signature of reentrance is observed 
both in $\chi_{ac}(T)$ \cite{visaniprl} and $\chi_{dc}(T)$, with the 
development of an additional paramagnetic susceptibility 
$\chi_{\mathrm para}(T)$, such that $\chi_{N}(T)=\chi_{\mathrm dia}(T)+\chi_{\mathrm para}(T)$.  

For sample 3AgNb, the susceptibility $\chi_{N}$ saturates 
below $T^{\mathrm sat}\approx 400\,\mu{\mathrm K}$, displaying a 
complete cancellation of the induced diamagnetic susceptibility in N, 
such that only the diamagnetism in S seems to remain.
For sample 5AgNb the susceptibility $\chi_{N}$ shows 
saturation below $T^{\mathrm sat}\approx 800\,\mu{\mathrm K}$ 
at a paramagnetic value $4\pi\chi_{N}^{\mathrm 
sat}\approx 1$, indicating a complete cancellation of the 
\textit{total} diamagnetic susceptibility in N plus S. 

The anomalously large magnitude of the paramagnetic reentrant 
susceptibility $\chi_{\mathrm para}$ at ultralow temperatures is rather 
intriguing.  This magnitude, and particularly its dependence on the 
sample size, is not explained by existing theories.  At this moment, 
it is not clear, if paramagnetic reentrance is (i) an intrinsic effect 
of mesoscopic NS proximity structures in the very low temperature 
limit or (ii) the result of two independent phenomena.  In the latter 
case the selection of the NS materials could be important.

A detailed inspection $\chi_{N}$ around its minimum reveals that, the 
reentrance temperature $T_{r}$ is decreased under a field 
$H_{dc}\approx 0.2\,{\mathrm Oe}$.  The ac and dc curves show reentrant 
temperatures $T_{r}=97\,{\mathrm mK}$ and $T_{r}^{H}=83\,{\mathrm mK}$ 
for sam-
\begin{figure}
\includegraphics[width=0.96\linewidth]{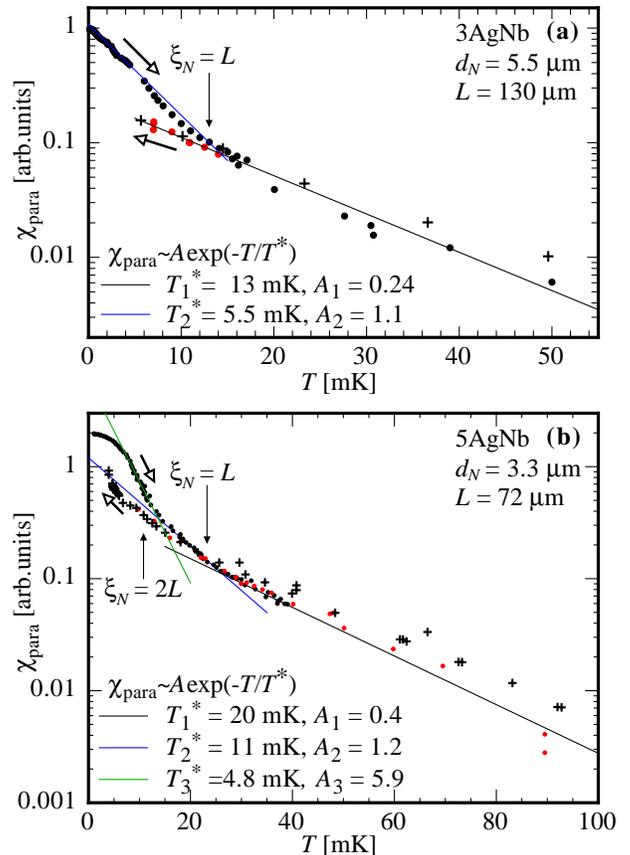}
		\caption{Reentrant paramagnetic susceptibility $\chi_{\mathrm 
		para}(T)$ below $T_{r}$.  For both samples two measurements of 
		$\chi_{dc}(T)$ ($\bullet$) are shown, and $\chi_{ac}(T)$ 
		($+$).  The thick arrows indicate the direction of temperature 
		changes.  The temperature scales are $\propto 1/L$.  The 
		vertical arrows indicate the temperatures at which the 
		condition $\xi_{N}(T)=nL$ is met.  }
        \protect\label{Xdc(T)log}
\end{figure}

\noindent ple 3AgNb ($T_{r}=149\,{\mathrm mK}$ and 
$T_{r}^{H}=113\,{\mathrm mK}$ for sample 5AgNb), respectively.

Neglecting the weak-field effect on $T_{r}$, $\chi_{dc}(T)$ matches 
$\chi_{ac}(T)$ for sample 3AgNb (5AgNb) between $15\,{\mathrm mK}$ 
($30\,{\mathrm mK}$) and $1\,{\mathrm K}$. This is rather noticeable, 
considering the strong nonlinearity in magnetic field discussed below.

Above $T\approx1\,{\mathrm 
K}$, $\chi_{dc}(T)$ deviates from $\chi_{ac}(T,H_{ac}\approx 0)$, due 
to the depression of the weak induced Andreev pair potential by finite 
fields.  Below $15\,{\mathrm mK}$ ($30\,{\mathrm mK}$), $\chi_{dc}(T)$ 
agrees with $\chi_{ac}(T)$ only on measurements done under cooling.  
Below $15\,{\mathrm mK}$ ($30\,{\mathrm mK}$) the warming up curve of 
$\chi_{dc}(T)$ lies above the cooling curves, corresponding to stronger 
paramagnetism adding to the full Meissner screening.

In our arrangement, samples are in rather good thermal contact with 
the Cu demagnetization stage and the Pt NMR thermometer.  Indeed, the 
measured thermal relaxation times at the lowest temperatures remain 
below $1000\,{\mathrm s}$.  Nevertheless, the susceptibility shows 
hysteresis. We have found the highest levels of paramagnetic 
reentrance only after allowing the NS system to remain at much below 
their saturation temperatures for long periods of time (one week or 
longer).

Fig.\ \ref{Xdc(T)log} shows the reentrant paramagnetic susceptibility 
$\chi_{\mathrm para}(T)$ below $T_{r}$.  The data exponentially 
increases as $\chi_{\mathrm para}(T)=A\exp{(-T/T^\ast)}$.  For sample 
3AgNb [Fig.\ \ref{Xdc(T)log}(a)], the prefactor $A_{1}=0.24$ and 
characteristic temperature $T_{1}^\ast=13\,{\mathrm mK}$ were obtained 
from $\chi_{ac}(T)$.  The cooling and warming $\chi_{dc}(T)$, above 
$\approx 13\,{\mathrm mK}$ reproduce well the behavior of 
$\chi_{\mathrm para}$ observed in $\chi_{ac}(T)$.  Near $\approx 
13\,{\mathrm mK}$, $\chi_{dc}$ displays a kink, leading to an 
approximate doubling of the logarithmic slope, with 
$T^\ast_{2}=5.5\,{\mathrm mK}\sim T^\ast_{1}/2$ and a prefactor 
$A_{2}=1.1$.

For sample 5AgNb [Fig.\ \ref{Xdc(T)log}(b)], the prefactor $A_{1}=0.4$ 
and characteristic temperature $T_{1}^\ast=20\,{\mathrm mK}$ were 
obtained from $\chi_{ac}(T)$.  Again the cooling and warming 
$\chi_{dc}(T)$ above $\approx 30\,{\mathrm mK}$ follow the 
$T$ behavior of $\chi_{ac}(T)$.  The curves show a kink in the 
susceptibility, displaying a doubling of the logarithmic slope, with 
$T^\ast_{2}=11\,{\mathrm mK}\sim T^\ast_{1}/2$ for the second line.  
In addition, around $\approx14\,{\mathrm mK}$, a second doubling of 
the logarithmic slope of $\chi_{dc}(T)$ occurs, with 
$T^\ast_{3}=4.8\,{\mathrm mK}\sim T^\ast_{2}/2$ for the third line.  

The coherence length of the Andreev pairs $\xi_{N}$ in Ag, obtained 
from our breakdown field measurements, is in agreement with the clean 
limit theory, $\xi_{N}=\hbar v_{F}/2\pi k_{B}T=1.69\,\mu{\mathrm 
m}/T({\mathrm K})$ \cite{bmueller1}.  At the temperature of the first 
kink in $\chi_{\mathrm para}$, $\xi_{N}(T)$ reaches approximately the 
value of a single wire's circumference $L=130\,\mu{\mathrm m}$ 
($L=72\,\mu{\mathrm m}$).  In Fig.\ \ref{Xdc(T)log} we have indicated 
by vertical arrows the temperature at which the equality 
$\xi_{N}(T)=L$ is met.  For sample 5AgNb, at the temperature of the 
second kink it is $\xi_{N}(T)=2L$.  The values of $T_{1}^\ast$, 
$T_{2}^\ast$, and $T_{3}^\ast$, as well as the position of the kinks, 
which are located approximately at $T_{1}^\ast$, $T_{2}^\ast$, give 
evidence for different levels of quantum coherence on the mesoscopic 
length scale $L$.  The temperatures $T^\ast$, which characterize the 
different levels as well as the kinks can be written as 
$T^\ast\approx\hbar v_{F}/2\pi k_{B}nL$, or in the equivalent form 
$\xi_{N}(T^\ast)=nL$, with $n=1,2,4$.  The reentrant paramagnetic 
susceptibility then is
\begin{equation}
	\label{eq1}
\chi_{\mathrm para}=A_{n} \exp{\left[-\frac{nL}{\xi_{N}(T)}\right]},\quad {\mathrm with}\, n=1,2,4\, .
\end{equation}

This characteristic behavior is not obtained by present theories. The 
correct theory of the paramagnetic reentrant effect should describe 
the suscptibility in accordance with Eq.\ \ref{eq1}. 

\begin{figure}
\includegraphics[width=0.96\linewidth]{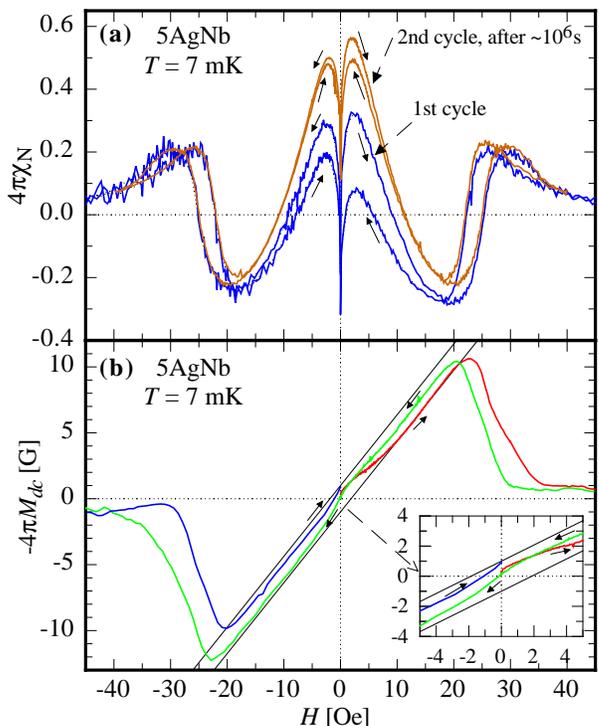}
		\caption{(a) Nonlinear ac susceptibility $\chi_{N}(H)$ 
		and (b) isothermal dc magnetization $M_{dc}(H)$ in full field 
		cycles, starting with positive values; 
		the black lines guide the eye.}
        \protect\label{XacMdc(H)nh}
\end{figure}

For a theoretical understanding of the origin of the reentrant effect, 
it is important to investigate the susceptibility under fields, as 
well as the magnetization.  In the following we discuss isothermal ac 
susceptibility measurements of sample 5AgNb, as a function of magnetic 
field [Figs.\ \ref{X(H,[T])nh} and \ref{XacMdc(H)nh}(a)].  The 
isothermal susceptibility $\chi_{N}(H)$ shows nonlinear behavior in 
the entire field regime.  At $7\,{\mathrm mK}$ two curves are shown, 
the first measurement directly after cooldown, and the second one 
\begin{figure}
\includegraphics[width=0.96\linewidth]{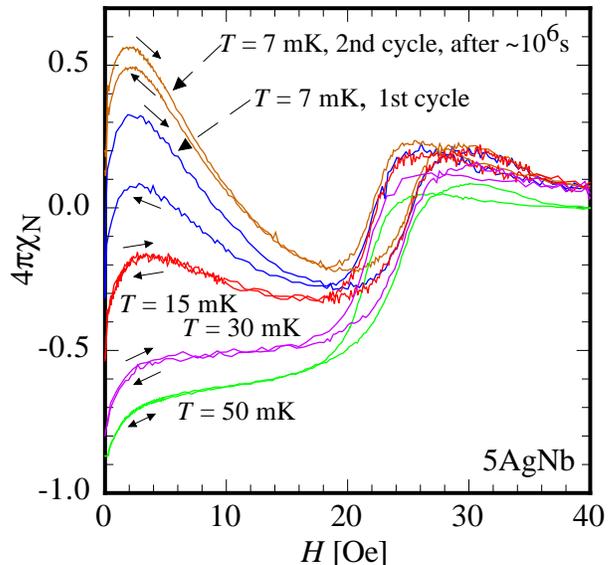}
		\caption{Isothermal ac-magnetic susceptibility $\chi_{N}(H)$, 
		below $50\,{\mathrm mK}$.
		The arrows indicate the direction of field changes.}
        \protect\label{X(H,[T])nh}
\end{figure}

\noindent $\sim 10^6\,{\mathrm s}$ after the first one.  The lower of the two 
curves at $7\,{\mathrm mK}$ shows pronounced hysteresis of the 
reentrant part at fields below $20\,{\mathrm Oe}$.  Near 20\,Oe the 
specimen screens the magnetic flux most effectively, before the well 
known magnetic breakdown transition occurs 
\cite{bmueller1,journlowtempphys,belzig:96,fauchere}.  For this 
sample, the transition is nearly temperature independent for 
$T<50\,{\mathrm mK}$.  However, the minimum susceptibility before the 
breakdown is far from complete screening.  At $T=7\,{\mathrm mK}$, it 
reaches only 30\,$\%$ of $-1/4\pi$, due to the paramagnetic 
contribution.

For very small dc magnetic fields, the magnetic susceptibility grows 
rather steeply.  At higher fields, that increase slows down, the 
susceptibility reaching its maximum value at about $2.5\,{\mathrm 
Oe}$, before turning towards less paramagnetic values.  As the 
temperature is increased, the curves show a less pronounced maximum at 
about $2.5\,{\mathrm Oe}$ and reduced hysteresis.  The detailed 
behavior at very low fields differs from the results of similar 
measurements, performed on a bigger wire bundle, reported in Ref.\ 
\cite{visaniprl}.  Possibly, the different geometrical arrangement of 
the wires in the bundles affect the average susceptibility.  Indeed, 
measurements on a single wire are desirable.

Cycling the magnetic field in both directions at $7\,{\mathrm mK}$ 
[Fig.\ \ref{XacMdc(H)nh}(a)], we observe two low-field peaks of 
$\chi_{N}(H)$, which are displaced symmetrically from zero (residual 
field $<2\,{\mathrm mOe}$).  However, the curves are not symmetric 
below $20\,{\mathrm Oe}$, displaying a reduction of hysteresis after 
each half-cycle.  Furthermore, after cooldown from above $50\,{\mathrm 
mK}$, cycling the field at $7\,{\mathrm mK}$, and waiting for $\sim 
10^6\,\mathrm{s}$, the low-field peak at about $2.5\,{\mathrm Oe}$ 
grows up.  This can be observed in the upper curves in Figs.\ 
\ref{X(H,[T])nh} and \ref{XacMdc(H)nh}(a).  After cycling the field 
and waiting, the whole system crosses to a more stable state, with 
more pronounced paramagnetic susceptibility.

Hysteresis and nonlinearity are also observed in dc magnetization 
curves, e.g. shown for a full cycle at $7\,{\mathrm mK}$ in Fig.\ 
\ref{XacMdc(H)nh}(b).  For the second half-cycle, nonlinearity becomes 
less pronounced, in accordance with our findings in the susceptibility 
curves.  The measured magnetization lies between two lines at fields 
below the breakdown transition.  At low fields one observes clearly 
the deviation of the magnetization from the induced Meissner 
screening.  At higher fields the curve asymptotically approaches the 
drawn line indicating linear Meissner-like behavior plus a constant 
\textit{paramagnetic} magnetization.  The $H=0$ intersections of the 
lines suggest a field-independent magnetization $4\pi M_{0}\approx 
1\,{\mathrm G}$.  It is interesting to notice that a spontaneous 
magnetization ot the same order was found in the theoretical model of 
Fauch\'ere \textit{et al}.  under the assumption of a negative average 
order parameter in N, $\Delta_{N}/k_{B}\approx-160\,{\mathrm mK}$ 
\cite{fauchere99}.  Moreover, some features of the paramagnetic 
reentrant effect, such as nonlinearity, hysteresis, and dissipation, 
are in qualitative agreemnet with their results \cite{fauchere99}.

At this point, a direct comparison of our experiment with theory 
\cite{bruder,fauchere99} is not possible.  The magnitude of 
$\chi_{\mathrm para}$ of order 1, the characteristic dependence of the 
magnetic susceptibility on $T$ and $L$, as described by Eq.\ 
\ref{eq1}, as well as its field dependence remain to be obtained by 
theory.  More theoretical work and also more experiments are 
necessary.

In summary, the paramagnetic reentrance phenomenon in AgNb cylinders 
of high purity is a nonlinear effect of anomalously strong magnitude.  
It shows strong deviations from induced Meissner screening in the 
low-temperature--low-field corner of the $H$-$T$ phase diagram. 
It displays dissipation \cite{visaniprl}, hysteresis, and long time 
relaxation behavior.

In the mesoscopic regime, the exponential temperature dependence of 
the magnetic susceptibility with characteristic temperature 
$T^\ast\approx\hbar v_{F}/2\pi k_{B}nL$ has the fingerprint of several 
levels of quantum coherence along integer multiples $n=1,2,4$ of the 
wire perimeter $L$.

Paramagnetic reentrance has also been observed in other NS materials 
\cite{visaniprl}.  In our most recent experiments we have found 
reentrant behavior in gold-niobium cylinders.  This has to be viewed 
in the light of expected superconductivity in Au below $T_{c}\approx 
200\,\mu{\mathrm K}$ \cite{hoyt}.  In consideration of this, the 
origin of this puzzling paramagnetism in mesoscopic NS cylinders is 
still an open question.

We wish to express our gratitude to R. Frassanito, M. Nider{\"o}st, 
and P. Visani for their invaluable contributions to earlier 
experimental work.  We thank the group at the ULT facility in Bayreuth 
for their help and support.  We acknowledge discussions with W. 
Belzig, G. Blatter, C. Bruder, A. Fauch\`ere, and Y. Imry.  We 
acknowledge partial support from the ``Schweizerischer Nationalfonds 
zur F{\"o}rderung der Wissenschaftlichen Forschung'' and the 
``Bundesamt f{\"u}r Bildung und Wissenschaft'' (EU Program ``Training 
and Mobility of Researchers'').

\end{multicols}


\begin{thebibliography}{00}
\bibitem{flukiger} {The initial sample preparation was done in the 
former lab of R. Fl{\"u}kiger at the Kernforschungszentrum Karlsruhe.
}
\end{thebibliography}
\end{document}